\documentclass[
10pt,nofootinbib, aps, pre
]{revtex4-1}

\usepackage{hyperref}
\usepackage{verbatim}
\usepackage{amsmath}
\usepackage{latexsym}
\usepackage{revsymb}
\usepackage{soul}
\usepackage{natbib}
\usepackage{amsfonts}
\usepackage{amsmath}
\usepackage{amssymb}
\usepackage{amsthm}
\usepackage{graphicx}
\usepackage{bm}

\def\beq{\begin{eqnarray}}
\def\eeq{\end{eqnarray}}
\newcommand{\be}{\begin{eqnarray} \begin{aligned}}
\newcommand{\ee}{\end{aligned} \end{eqnarray} }

\usepackage{amsfonts}

\def\const{\operatorname{const}}

\begin{document}

\title{Bound states in weakly deformed waveguides: numerical vs analytical results}

\author{Paolo Amore}
\email{paolo.amore@gmail.com}
\affiliation{Facultad de Ciencias, CUICBAS, Universidad de Colima, \\
Bernal D\'{\i}az del Castillo 340, Colima, Colima, Mexico}
\author{John P. Boyd}
\email{jpboyd@umich.edu}
\affiliation{Department of Atmospheric, Oceanic \& Space Science  \\
University of Michigan, 2455 Hayward Avenue, Ann Arbor MI 48109}
\author{Francisco M. Fern\'andez}
\email{fernande@quimica.unlp.edu.ar}
\affiliation{INIFTA (UNLP, CONICET), Division Quimica Teorica, \\
Blvd. 113 S/N, Sucursal 4, Casilla de Correo 16, 1900 La Plata, Argentina}
\author{Martin Jacobo}
\affiliation{Facultad de Ciencias, Universidad de Colima, \\
Bernal D\'{\i}az del Castillo 340, Colima, Colima, Mexico}
\author{Petr Zhevandrov}
\email{pzhevand@gmail.com}
\affiliation{Facultad de Ciencias F\'isico-Matem\'aticas, \\
Universidad Michoacana de San Nicol\'as de Hidalgo,
Ciudad Universitaria, 58030 Morelia, Michoac\'an, Mexico}

\begin{abstract}
We have studied the emergence of bound states in weakly deformed and/or heterogeneous waveguides, 
comparing the analytical predictions obtained using a recently developed perturbative method, with precise 
numerical results, for different configurations (a homogeneous asymmetric waveguide, a heterogenous
asymmetric waveguide and a homogeneous broken-strip). 
In all the examples considered in this paper we have found excellent agreement between analytical 
and numerical results, thus providing a numerical verification of the analytical approach.
\end{abstract}

\maketitle


\section{Introduction}
\label{sec:Intro}
The appearance of trapped modes (bound states) in open geometries under perturbations has attracted a lot of attention in both physical and mathematical literature in the recent past (see, e.g., the books \cite{Lond, Hurt, ExH}, where an interested reader can find a rather complete bibliography). The classical example is the appearance of a bound state for the one-dimensional Schr\"odinger operator perturbed by a small potential well \cite{Sim76}. In this situation the unperturbed problem possesses a purely continuous spectrum corresponding to plane waves $\exp\{ikx\}$ with the energy $E=k^2$ so that the continuous spectrum occupies the positive ray $0\leq E<\infty$. Under a perturbation by a potential well $\epsilon V(x)$ with $\epsilon\to+0$ and a smooth and compactly supported function $V$ such that $\int V(x)\,dx<0$, the threshold $E=0$ gives rise to a bound state whose energy $E_\epsilon$ is located to the left of the continuous spectrum close to the threshold, $E_\epsilon=-\const\epsilon^2$, where the constant is proportional to the square of the area above the graph of $V(x)$ (see \cite{Sim76} for details).

This mechanism of the generation of bound states by the threshold of the continuous spectrum seems to be quite generic and has analogs in many situations of physical interest (see the books cited above). For example, Bulla et al. \cite{Bulla97} discovered that a similar phenomenon occurs in a slightly deformed waveguide described by the Laplace operator with Dirichlet boundary conditions on the walls. They have shown that under certain perturbations (which enlarge the waveguide) the threshold of the continuous spectrum gives rise to an eigenvalue to the left of it whose distance to the continuous spectrum is analytic in the perturbation parameter in a neighborhood of the origin, and also gave an explicit formula for the leading term in the expansion of the eigenvalue in the Taylor series with respect to this parameter. They used the so-called Birman-Schwinger technique to obtain these results. Subsequently, many other approaches to the problems of this kind were developed (see, e.g., \cite{Naz, Borisov01, ExH}). For the most part, these techniques provide approximate formulas for the eigenvalues up to a certain power of the perturbative parameter, but of course it is impossible to say whether the approximate formulas in fact approximate the eigenvalues for a given fixed value of the parameter. One of the goals of the present paper is the investigation of this question for several examples from the waveguide theory by means of the comparison of the precise numerical results obtained by means of the collocation method with the variational estimates and with the first nonvanishing terms provided by the theoretical formulas of different perturbative approaches.

The principal difficulty of the problems under consideration  is that the nonperturbed problem does not have eigenvalues, so that the standard regular perturbation theory is not applicable. Recently, still another perturbative approach, which extends a method previously developed in \cite{Gat93}, was proposed in \cite{Amore15, Amore16}. It has the advantage of using an auxiliary ``unperturbed'' problem which does possess an eigenvalue and is exactly solvable, so that a standard perturbation procedure can be used in order to construct the corrections up to any order. The comparison of the previously known theoretical results, as well as the precise numerical calculations, with this new perturbative approach  constitutes the second goal of the paper. 

We note that the approach of \cite{Amore15, Amore16} is equally applicable to weakly deformed and/or weakly heterogeneous waveguides. Therefore, we have chosen the following three examples: a) an asymmetrically deformed waveguide (this is exactly the case considered in \cite{Bulla97}); b) an asymmetrically deformed waveguide  with a localized heterogeneity (we are not aware of the existence of any results in this case in the literature); and c) a broken strip (investigated by  different methods in, e.g., \cite{Avishai91, Granot02}).

The paper is organized as follows: in Section II we summarize the results of \cite{Amore15, Amore16}, in Section III we present the comparison of the results of different approaches to the three examples mentioned above, and in Section IV we give our conclusions.

\section{The method}
\label{sec:Method}

The problem of calculating the emergence of trapped states in infinite, slightly heterogeneous waveguides has been recently
considered in two papers, refs.~\cite{Amore15,Amore16}, where exact perturbative formulas which use the density
inhomogeneity as a perturbation parameter were obtained. This approach extends a method previously developed by Gat and Rosenstein \cite{Gat93},
for calculating the binding energy of threshold bound states.

In particular ref.~\cite{Amore15} contains the calculation up to third order,
while ref.~\cite{Amore16} extends the calculation to fourth order.
These formulas have been tested there on two exactly solvable models, reproducing the exact results up to
fourth order.

Although the examples considered there were limited to the case of heterogeneous straight waveguides, the
perturbative expressions apply as well to the case of homogeneous, slightly deformed waveguides and to
the more general case of slightly heterogeneous and slightly deformed waveguides. The present paper focuses on the
application to these last two cases. Incidentally, while the effect of small deformations on
infinite and homogeneous waveguides has been studied before by different authors and using different techniques, the
effect of weak heterogeneities on infinite (either straight or deformed) waveguides is much less known.
In this paper we perform a comparison between the theoretical predictions of the formulas obtained in refs.~\cite{Amore15,Amore16}
with the numerical, precise results, for different models.

We refer the reader interested in the details of the perturbative expansion to refs.~\cite{Amore15,Amore16} and here
limit ourselves to report the general formulas for the perturbative corrections to the energy of the fundamental mode
of a heterogeneous waveguide, up to fourth order, that read
\begin{eqnarray}
E_0^{(1)} &=& 0  \\
E_0^{(2)} &=& - \frac{\pi^2}{b^2} \Delta_2^2 \\
E_0^{(3)} &=& - 2 \frac{\pi^2}{b^2} \Delta_2 (\Lambda_1 - \Delta_3) \\
E_0^{(4)} &=&   -\frac{\pi^2}{b^2}  \left[ -2\Delta_2^4 - \Delta_2^2 \Delta_4 + 2 \Delta_2 \Delta_5 + \Delta_3^2 - 2 \Lambda_2 - \Delta_3 \Lambda_1
+ 2 \Delta_2 \Lambda_3 + \Lambda_1^2 \right]
\end{eqnarray}
where we use the definitions introduced in ref.~\cite{Amore16}
\begin{eqnarray}
\Delta_1 &\equiv&  \frac{\pi}{b^2} \int dxdy \sigma(x,y) \nonumber \\
\Delta_2 &\equiv&  \frac{\pi}{b^2} \int dxdy \sigma(x,y) \cos^2 \frac{\pi y}{b} \nonumber \\
\Delta_3 &\equiv& \frac{\pi^3}{b^5} \int dx_1dy_1 \int dx_2dy_2 \sigma(x_1,y_1) \sigma(x_2,y_2) |x_1-x_2|
\cos^2 \frac{\pi y_1}{b} \cos^2 \frac{\pi y_2}{b} \nonumber \\
\Delta_4 &\equiv& \frac{\pi^4}{b^6} \int dx_1dy_1 \int dx_2dy_2 \sigma(x_1,y_1) \sigma(x_2,y_2) x_1 (2x_2-x_1)
\cos^2 \frac{\pi y_1}{b} \cos^2 \frac{\pi y_2}{b} \nonumber \\
\Delta_5 &\equiv& \frac{\pi^5}{b^8} \int dx_1dy_1 \int dx_2dy_2 \int dx_3dy_3 \sigma(x_1,y_1) \sigma(x_2,y_2)\sigma(x_3,y_3) |x_1-x_2|  |x_2-x_3| \nonumber \\
&\times& \cos^2 \frac{\pi y_1}{b} \cos^2 \frac{\pi y_2}{b} \cos^2 \frac{\pi y_3}{b}  \nonumber \\
\Lambda_1 &\equiv& \frac{\pi^3}{b^4} \int dx_1dy_1 \int dx_2dy_2 \sigma(x_1,y_1) \sigma(x_2,y_2)
\cos \frac{\pi y_1}{b} \cos \frac{\pi y_2}{b} g_2^{(0,0)}(x_1,y_1,x_2,y_2)  \nonumber \\
\Lambda_2 &\equiv& \frac{\pi^6}{b^9} \int dx_1dy_1 \int dx_2dy_2 \int dx_3dy_3 \int dx_4dy_4 \sigma(x_1,y_1) \sigma(x_2,y_2) \sigma(x_3,y_3) \sigma(x_4,y_4)
|x_1-x_3| \nonumber \\
&\times& \cos \frac{\pi y_1}{b} \cos \frac{\pi y_2}{b} \cos^2 \frac{\pi y_3}{b} \cos^2 \frac{\pi y_4}{b}  g_2^{(0,0)}(x_1,y_1,x_2,y_2)  \nonumber \\
\Lambda_3 &\equiv& \frac{\pi^5}{b^6} \int dx_1dy_1 \int dx_2dy_2 \int dx_3dy_3 \sigma(x_1,y_1) \sigma(x_2,y_2) \sigma(x_3,y_3)
 \nonumber \\
&\times& \cos \frac{\pi y_1}{b}  \cos \frac{\pi y_3}{b}   g_2^{(0,0)}(x_1,y_1,x_2,y_2) g_2^{(0,0)}(x_2,y_2,x_3,y_3)   \nonumber
\end{eqnarray}

The correction to the energy of the fundamental mode up to fourth order can be cast in  the form
\begin{eqnarray}
\Delta E_0 &\approx& E_0^{(2)}  + E_0^{(3)} + E_0^{(4)} = - \frac{\pi^2}{b^2} \left\{  \left( \Delta_2 + (\Lambda_1- \Delta_3)^2 \right)^2 +  \Gamma \right\} \ ,
\label{EQ_DE}
\end{eqnarray}
where
\begin{eqnarray}
\Gamma \equiv \left[ - 2 \Delta_2^4 + \Delta_2 \Delta_3 - \Delta_2^2 \Delta_4 + 2 \Delta_2 \Delta5 - \Delta_3 \Lambda_1 -2 \Lambda_2 +2 \Delta_2 \Lambda_3\right] \ .
\end{eqnarray}

Eq.~(\ref{EQ_DE}) will be applied in the next section to different waveguides.

\section{Application to deformed waveguides}

The perturbative formulas obtained in Refs.~\cite{Amore15,Amore16} apply to the general case of heterogeneous
and deformed waveguides, although the applications considered in those papers concerned only slightly heterogeneous
straight waveguides.

An appropriate conformal map $w \equiv u + i v = F(z)$ can map an infinite strip, $y \in (b_-,b_+)$, with $b_- < b_+$, 
and  $- \infty < x<\infty$, onto a deformed waveguide, $u \in (-\infty, \infty )$ and $f_-(u) < v < f_+(u)$.
Here $f_\pm(u)$ are the upper and lower borders of the deformed strip, over which Dirichlet boundary conditions
are assumed.

Suppose that one has to solve the Helmholtz equation for the deformed strip, assumed to be heterogeneous, and
with a physical density varying at each point, $\rho(u,v) > 0$. We also require that the density variations are 
small and localized  around one (or more points) internal to the domain or equivalently that $\rho(u,v)$ tends 
sufficiently rapidly to a constant value $\rho_0$ as $|u|\to\infty$.

One then has to solve the eigenvalue equation
\begin{eqnarray}
-\Delta_{u,v} \phi(u,v) = E \rho(u,v)  \phi(u,v)
\label{eq:Helmholtz1}
\end{eqnarray}
with $\phi(u,f_\pm(u))=0$ and $\Delta_{u,v} \equiv \frac{\partial^2}{\partial u^2} + \frac{\partial^2}{\partial v^2}$.

If we map the deformed strip back to the straight waveguide, the equation above transforms into
\begin{eqnarray}
-\Delta_{x,y} \psi(x,y) = E \Sigma(x,y) \rho(u(x,y),v(x,y))  \psi(x,y)
\label{eq:Helmholtz2}
\end{eqnarray}
where $\Sigma(x,y) =  \left| \frac{dF}{dz}\right|^2$ (we will refer to it as to the ``conformal density'')
and $\psi(x,b_\pm ) =0$.

From a physical point of view, eq.~(\ref{eq:Helmholtz1}) can be interpreted
as the Helmholtz equation for a straight waveguide with a physical density $\tilde{\rho}(x,y) \equiv
\Sigma(x,y) \rho(u(x,y),v(x,y))$.

Under the assumptions of small deformations and of weak heterogeneity,  one can write
\begin{equation}
\tilde{\rho}(x,y) = \rho_0 +  \tilde{\sigma}(x,y)
\end{equation}
where $|\tilde{\sigma}(x,y)| \ll 1$ and $\lim_{|x| \rightarrow \infty}\tilde{\sigma}(x,y) = 0$.

In this case eq.~(\ref{eq:Helmholtz2}) has precisely the form  discussed in Refs.~\cite{Amore15,Amore16}
and one can straightforwardly apply the perturbative formulas obtained in Refs.~\cite{Amore15,Amore16} to calculate
the corrections to the lowest eigenvalue, for a waveguide which is both deformed and
heterogeneous~\footnote{Finding the conformal map which sends a given deformed waveguide
into a straight waveguide may still be a difficult challenge. We are not concerned with this
issue here. }.

In the following we will examine three examples: an asymmetric homogeneous waveguide,  with a local
enlargement, an asymmetric heterogeneous waveguide with a local narrowing, and a slightly broken strip, with a
homogeneous density.

\subsection{Asymmetrically deformed waveguide}
\label{sub_adw}

Bulla and collaborators \cite{Bulla97} have considered the waveguide on the domain
\begin{eqnarray}
\Omega_\lambda = \left\{ (x,y) \in \mathbb{R}^2 | 0 < y <  1+\lambda f(x) \right\}
\end{eqnarray}
and found that the fundamental mode of the Laplacian on this domain, for Dirichlet boundary conditions
at the border, behaves as
\begin{eqnarray}
E(\lambda) = \pi^2 - \pi^4 \lambda^2 \left( \int_{\mathbb{R}} f(x) dx \right)^2 + O(\lambda^3)
\label{EQ_Bulla}
\end{eqnarray}

\begin{figure}
\begin{center}
\bigskip\bigskip\bigskip
\includegraphics[width=12cm]{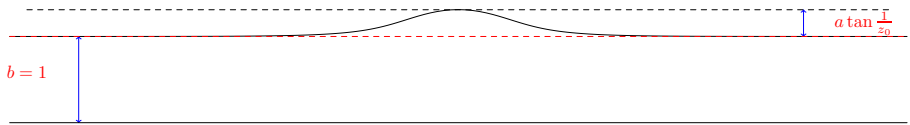}
\caption{Asymmetric waveguide}
\label{Figure_1}
\end{center}
\end{figure}

Since the formulas of Refs.~\citep{Amore15,Amore16} apply to this domain as well, we consider the
conformal map:
\begin{eqnarray}
F(z) = z+a \tanh \left(\frac{z}{z_0}\right) \label{eq:map}
\end{eqnarray}
which transforms a straight waveguide of unit width into a waveguide of the kind considered by
Bulla et al. The waveguide obtained using this map for $a=1/5$ and $z_0=1$ is displayed in Fig.~\ref{Figure_1}.
Observe that $F(z)$ has simple poles at $z = i (2k+1)\pi z_0 /2$ with $k=0,\pm 1,\pm 2, \dots$.

The lower side of the waveguide is not deformed, whereas the upper side is deformed to the parametric
curve
\begin{eqnarray}
\left\{ 
\begin{array}{ccc}
u(x) & = & x+\frac{a \sinh \left(\frac{2x}{z_0}\right)}{\cos\left(\frac{2}{z_0}\right)+\cosh \left(\frac{2 x}{z_0}\right)} \\
v(x) & = & 1+\frac{a \sin \left(\frac{2}{z_0}\right)}{\cos\left(\frac{2}{z_0}\right)+\cosh \left(\frac{2 x}{z_0}\right)}  \\
\end{array}
   \right.
\end{eqnarray}
with $x \in (-\infty,\infty)$.

In this case the conformal density is given by
\begin{eqnarray}
\Sigma(x,y) &=& 1 + \sigma(x,y)
\end{eqnarray}
where
\begin{eqnarray}
\sigma(x,y) &=& \frac{4 a \left(\cosh \left(\frac{2 x}{z_0}\right)
\cos\left(\frac{2 y}{z_0}\right)+1\right)}{z_0\left(\cosh \left(\frac{2 x}{z_0}\right)
+\cos\left(\frac{2 y}{z_0}\right)\right)^2}
+ \frac{4 a^2}{z_0^2 \left(\cosh \left(\frac{2x}{z_0}\right)+\cos \left(\frac{2y}{z_0}\right)\right)^2}
\end{eqnarray}

The area corresponding to the enlargement (i.e.  to $1 < y < 1+a \tan 1/z_0$)
is obtained with the formula\footnote{Observe that in this case we are not allowed
to calculate the enlargement with the formula $\int_{-\infty}^\infty dx \int_0^1 dy (\Sigma(x,y)-1)$,
since both integrals $\int dxdy \Sigma$ and $\int dxdy 1$ are divergent. }
\begin{eqnarray}
\delta\mathcal{A} &=& \int_{-\infty}^\infty f(u) du =
\int_{-\infty}^\infty \frac{a \sin
   \left(\frac{2}{{z_0}}\right)}{\cosh
   \left(\frac{2 x}{{z_0}}\right)+\cos
   \left(\frac{2}{{z_0}}\right)} dx + O(a^2) \nonumber \\
&=& 2 a + O(a^2)
\end{eqnarray}

For $\delta\mathcal{A} >0$  one can apply the formula of Bulla et al. and the
lowest eigenvalue of the deformed waveguide reads
\begin{eqnarray}
E \approx \pi^2 - 4 \pi^4 a^2 +  O(a^3)
\end{eqnarray}

We now calculate the dominant behavior of lowest eigenvalue of the waveguide using
the expression for the second order contribution of Ref.~\cite{Amore15};
in this case we have
\begin{eqnarray}
\Delta_2 &=& \pi \int_{-\infty}^{\infty} dx  \int_0^1 dy \sigma(x,y) \sin^2 (\pi y)) =  2 \pi a + O(a^2)
\end{eqnarray}
and obtain
\begin{eqnarray}
E_0^{(2)} = - \pi^2 \Delta_2^2 = -4 \pi ^4 a^2 + O(a^3)
\end{eqnarray}
that agrees with the result obtained using the formula of Bulla et al., as it should.

To assess the quality of the perturbative estimates we have also obtained variational bounds
on the energy of the fundamental mode, using the ansatz
\begin{eqnarray}
\Psi(x,y) = \sqrt{2\gamma} \sin (\pi y) \ e^{-\gamma \sqrt{\delta^2+x^2}}  \
\left( 1 + \eta \frac{1-y}{1+\epsilon^2 x^2} \right)
\end{eqnarray}
where $\gamma$, $\delta$,$\eta$ and $\epsilon$ are variational parameters.

The variational bound is
\begin{eqnarray}
E \leq E_0^{(var)} = \frac{\langle \Psi | -\nabla^2 | \Psi  \rangle}{\langle\Psi | \Sigma | \Psi \rangle}
\end{eqnarray}

\begin{figure}
\begin{center}
\bigskip\bigskip\bigskip
\includegraphics[width=4.5cm]{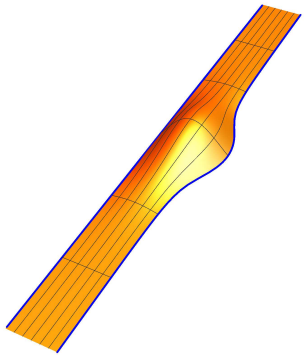}
\caption{Variational wave function for the asymmetric waveguide for $a=1/3$ and $z_0=1$.}
\label{Figure_2}
\end{center}
\end{figure}

The wavefunction in Fig.\ref{Figure_2} was obtained using the variational ansatz above,
for the case of a waveguide with $a=1/3$ and $z_0 =1$, and minimizing the
Rayleigh quotient with respect to the variational parameters
(this case is nonperturbative and the perturbative formulas cannot be applied).

\begin{table}
\caption{\label{tab:table2} Comparison between the variational bounds, the numerical results
obtained using a collocation approach and the perturbative corrections
for the asymmetric waveguide corresponding to the map (\ref{eq:map}) with $z_0=1$ for different values of $a$.
Underlined digits in the third column are believed to be stable.}
\begin{center}
\begin{tabular}{|c|l|l|l|}
\hline
$a$  & $E_0^{(var)} - \pi^2$ & $E_0^{(collocation)} - \pi^2$ & $E_{0}^{(2)}$   \\
\hline
$10^{-5}$ & $-3.89390 \times 10^{-8}$ &  $-\underline{3.895} \times 10^{-8}$ & $-3.89636 \times 10^{-8}$ \\
$10^{-4}$ & $-3.88415 \times 10^{-6}$ & $-\underline{3.8846} \times 10^{-6}$ & $-3.89636 \times 10^{-6}$ \\
$10^{-3}$ & $-3.78211 \times 10^{-4}$ & $-\underline{3.7832}2 \times 10^{-4}$ & $-3.89636 \times 10^{-4}$ \\
$10^{-2}$ & $-3.03276 \times 10^{-2}$ &  $-\underline{3.03339}6 \times 10^{-2}$ & $-3.89636 \times 10^{-2}$ \\
$10^{-1}$ & $-1.08811$ & $-\underline{1.088648}7$   & $-3.89636$ \\
\hline
\end{tabular}
\end{center}
\end{table}

Additionally, we calculated the lowest eigenvalue by a collocation (“pseudospectra’”, “discrete ordinates”)  method. 
As a check, we employed both the rational Chebyshev basis \cite{Boyd87} and the sinh-Fourier (Cloot-Weideman) basis \cite{Cloot90}.  
Both imply a domain which is a strip of uniform width, conformally mapped from the asymmetric waveguide. Conformal mapping has fallen 
out of favor as a grid generation scheme for numerical PDE solvers because of uncontrollable and sometimes extreme nonuniformity 
(this is the "Geneva effect", so named because like the diplomatic talks so frequent in that city, conformal mapping of a highly 
deformed region can bring distant groups together).  Here, the conformal map is a small perturbation of the identity transformation 
and no such difficulties arise. Non-conformal coordinate changes  introduce a lot of metric factors into the transformed partial 
differential equation in contrast to the single metric factor $1 + \sigma(x,y)$ displayed in (8).”

In Table \ref{tab:table2} we compare the variational bounds obtained using the ansatz above with
results obtained using collocation and with the  second order perturbative estimate,
for different values of $a$, keeping $z_0 = 1$. For $a \rightarrow 0$ the results obtained
with the different methods (variational, collocation and perturbative) are very close.
Note that, for very small values of $a$, the wave function decays extremely slowly as
$|x| \rightarrow \infty$ and the application of the collocation method becomes more challenging.


\subsection{Asymmetrically deformed waveguide with a localized heterogeneity}
\label{sub_adw2}

We consider now the case of a waveguide studied in the subsection \ref{sub_adw}, in presence of a localized
inhomogeneity, represented by the density
\begin{eqnarray}
\rho(u,v) = 1 + \rho_0 e^{-\zeta u^2}
\end{eqnarray}
where it is assumed that $|\rho_0| \ll 1$ and $\zeta>0$.

After applying the conformal map, one converts the original problem to an equivalent problem with a density
\begin{eqnarray}
\tilde{\rho}(x,y) 
&\approx& 1 +  \frac{4 a (1+\cos (2 y) \cosh (2 x))}{(\cos (2 y)+\cosh (2x))^2}+e^{-x^2 \zeta } \rho _0
\end{eqnarray}
where we have assumed in the last equation that $|a| \approx |\rho_0| \ll 1$ and neglected the contribution depending on $a\rho_0$.

In this case, one can apply the formalism of Refs.~\cite{Amore15,Amore16}, to calculate the leading order correction to the lowest
eigenvalue using the density $\tilde{\rho}$:
\begin{eqnarray}
\Delta_2 = \pi \int_{-\infty}^\infty dx \int_{0}^1 dy \tilde{\rho}(x,y) \sin^2 \pi y  \approx \pi  \left(2 a+\frac{\sqrt{\pi } \rho _0}{2 \sqrt{\zeta }}\right)
\end{eqnarray}

\begin{figure}
\begin{center}
\bigskip\bigskip\bigskip
\includegraphics[width=12cm]{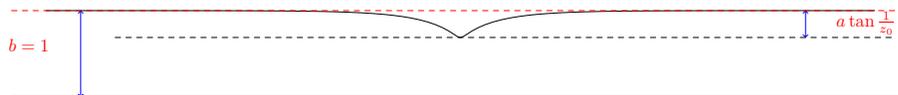}
\caption{Asymmetric waveguide with $a<0$.}
\label{Figure_3}
\end{center}
\end{figure}

As discussed in Ref.~\cite{Amore15}, the condition $\Delta_2 > 0$ implies the existence of a bound state; in particular, when $a<0$ and $\Delta_2>0$, 
one is considering a waveguide with a small entering deformation of the upper border and a weak inhomogeneity, which  however is sufficient to provide 
binding. Interestingly, an arbitrarily small $\rho_0$ will still provide binding if the inhomogeneity  is distributed over a sufficiently large region 
(i.e. if $\zeta$ is sufficiently small). The case of a waveguide with $a<0$ is represented in Fig.~\ref{Figure_3}.

The leading correction to the lowest eigenvalue is now
\begin{eqnarray}
E_0^{(2)} = - \pi^2 \Delta_2^2 \approx -\pi^4  \left(2 a+\frac{\sqrt{\pi } \rho _0}{2 \sqrt{\zeta }}\right)^2
\label{EQ_PT2}
\end{eqnarray}

Note that the excess mass distributed on the waveguide can be easily calculated
\begin{eqnarray}
\delta M =  \int du dv \rho(u,v) \approx \sqrt{\frac{\pi}{\zeta}} \rho_0
\end{eqnarray}
and the energy in eq.~(\ref{EQ_PT2}) can thus be written in a form
similar to eq.~(\ref{EQ_Bulla}) as
\begin{eqnarray}
E_0^{(2)}  \approx -\pi^4 \left(\delta\mathcal{A} + \frac{\delta M}{2}\right)^2
\end{eqnarray}

\begin{figure}
\begin{center}
\bigskip\bigskip\bigskip
\includegraphics[width=8cm]{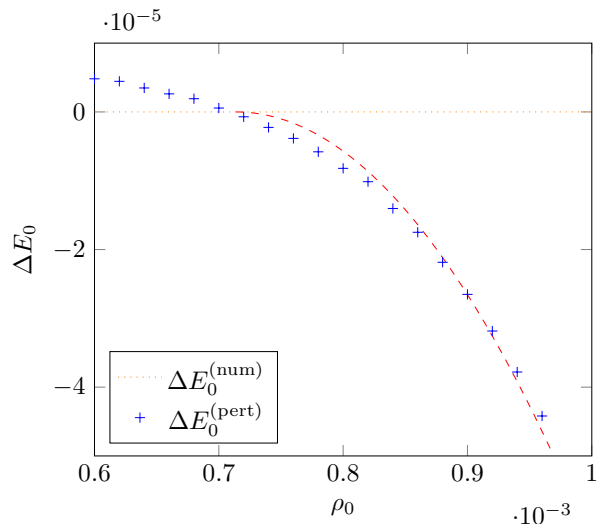}
\caption{$\Delta E_0$ for an asymmetric waveguide with $a=-0.001$ and $\zeta = 0.1$, as a function of $\rho_0$. The dashed line
corresponds to the theoretical prediction of Eq.(\ref{EQ_PT2}); the symbols ``+'' correspond to the numerical values obtained using
collocation with mapped Chebyshev functions (a set with $101$ functions along the horizontal direction and $11$ along the vertical
direction).}
\label{Figure_4}
\end{center}
\end{figure}

The accuracy of this formula has been verified in Fig.~\ref{Figure_4} for an asymmetric waveguide with $a=-0.001$ and $\zeta = 0.1$,
calculating the energy shift $\Delta E_0 = E_0 -\pi^2$ as a function of $\rho_0$. The dashed line is the theoretical prediction
of our second order formula, whereas the ``+''-symbols correspond to the numerical result obtained using mapped Chebyshev functions
(the scale $L=9$ has been used in all calculations and a set of $101$ functions along $x$ and $11$ along $y$ has been used).
Observe that the critical value of $\rho_0$ is quite similar in both calculations; there is however
a mild discrepancy between numerical and theoretical values for $\rho_0$ sufficiently close to the critical value.
The explanation of this discrepancy is straightforward: as $\rho_0$ approaches the critical value, the wave function
decays slower and slower for $|x| \rightarrow \infty$ and therefore the numerical calculation should use larger sets of functions
to mantain the same accuracy. On the other hand, for $\rho_0$ sufficiently large (not shown in the figure), one also expects a discrepancy between the two curves, due to the non-perturbative nature of the solution.

\subsection{Broken strip}
\begin{figure}
\begin{center}
\bigskip\bigskip\bigskip
\includegraphics[width=10cm]{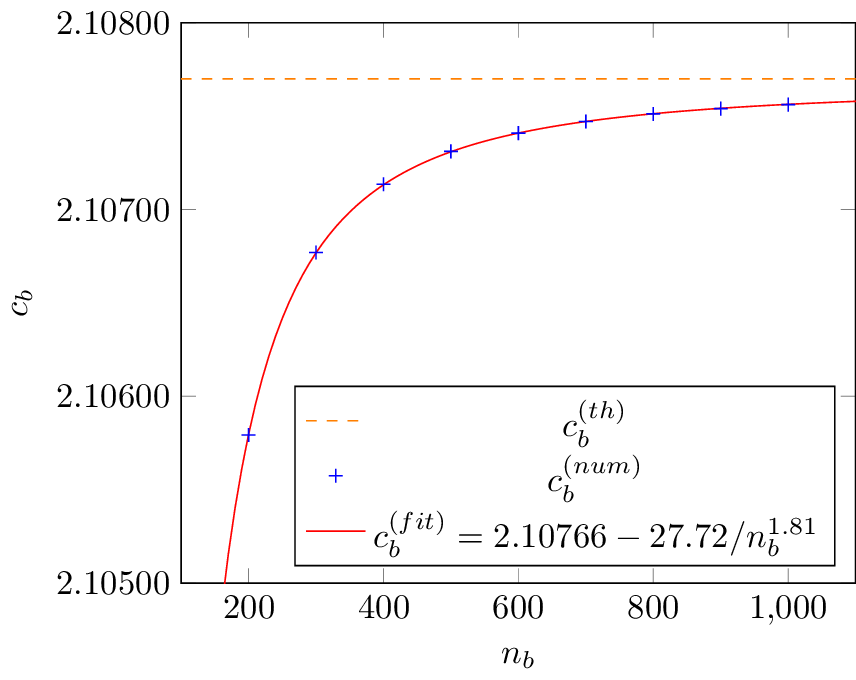}
\caption{Comparison between the numerical and theoretical estimates of $c_b$.}
\label{Figure_6}
\end{center}
\end{figure}

Our third example is a broken strip, i.e. an infinite waveguide, of constant unit
width, where the two semi-infinite arms form an angle $\alpha$, as displayed in Fig.~\ref{Figure_6}.
This problem has been studied in a series of papers,
refs.~\cite{Avishai91,Carini92,Carini93, Exner95,Granot02,Levin04,Sadurni10, Bittner13}.

In particular Avishai et al., ref.~\cite{Avishai91}, Duclos and Exner, ref.\cite{Exner95}, and
Granot ref.~\cite{Granot02}, have studied the case of a weak bending, corresponding to
the limit $\phi \rightarrow 0$. Avishai and collaborators found that in this regime the energy of the bound state
behaves as
\begin{eqnarray}
E_0  \approx \pi^2 - c_b \phi^4
\end{eqnarray}
with $c_b \approx 2.10$.

The regime corresponding to sharp bendings has been studied recently in Refs.~\cite{Sadurni10, Bittner13},
using an effective potential approach, and  tested experimentally using electromagnetic waveguides.
The theoretical approach of Refs.~\cite{Sadurni10, Bittner13} relies on the use of a conformal map, which
transforms the broken strip into an infinite straight waveguide, and the original Helmholtz equation into
a Schr\"odinger-like equation, with an effective potential (see eq.~(19) of Ref.~\cite{Bittner13}).

The conformal map considered in Refs.~\cite{Sadurni10, Bittner13} is given by
\begin{eqnarray}
F(z) = \frac{1}{\pi } \ B\left(\sin ^2\left(\frac{\pi  z}{2}\right),\frac{1}{2}-\frac{\phi}{\pi },\frac{1}{2}+\frac{\phi }{\pi }\right);
\end{eqnarray}
where 
\begin{eqnarray}
B(x,p,q) \equiv \int_0^x u^{p-1} (1-u)^{q-1} du \nonumber
\end{eqnarray}
is the incomplete beta function. $F(z)$ maps the unit strip $x \in (0,1)$ onto a broken strip of unit width, with the arms forming an angle $\alpha = \pi-2\phi$. 

Using this conformal map we may transform the original equation into the form of Eq.~(\ref{eq:Helmholtz2})
where
\begin{eqnarray}
\Sigma(x,y) &=&  \left| \frac{dF}{dz}\right|^2
 =  \left(1+\frac{2}{\sec (\pi  x) \cosh (\pi  y)-1}\right)^{\frac{2 \phi }{\pi }} \equiv 1+ \sigma(x,y)
\end{eqnarray}
is the "conformal density" and  $x \in (0,1)$ and $|y| < \infty$.
Note that in this case, the waveguide is arranged vertically, and therefore the perturbative expressions
of Refs.\cite{Amore15,Amore16}  should be adapted.

\begin{figure}
\begin{center}
\bigskip\bigskip\bigskip
\includegraphics[width=8cm]{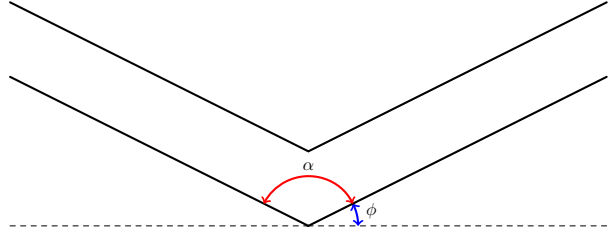}
\caption{Broken strip.}
\label{Figure_5}
\end{center}
\end{figure}

For small bendings, we may expand this density about $\phi = 0$ and obtain
\begin{eqnarray}
\Sigma(x,y) &\approx& 1 + \frac{2}{\pi } \phi \log \left(1 + \frac{2}{\sec (\pi  x) \cosh (\pi y)-1}\right)
+\frac{2}{\pi ^2} \phi^2 \log ^2\left(1+\frac{2}{\sec (\pi  x) \cosh (\pi y)-1}\right) \nonumber \\
&+& \frac{4}{3 \pi^3} \phi^3 \log ^3\left(1+\frac{2}{\sec (\pi  x) \cosh (\pi y)-1}\right) + \dots
\end{eqnarray}
where the density perturbation $\sigma$ can now be easily read off this expression.

Under these conditions, we are in the position of applying directly the approach developed in ref.~\cite{Amore15,Amore16}. In particular, the second order correction to the  perturbative expansion
reads
\begin{eqnarray}
E_0^{(2)} &=& - \pi^2 \Delta_2^2
\end{eqnarray}
where
\begin{eqnarray}
\Delta_2 &\equiv& \pi \int_0^1 dx \int_{-\infty}^{\infty} dy \ \sigma(x,y) \sin^2 (\pi x) \nonumber \\
&=&  \pi \frac{2}{\pi ^2} \phi^2  \int_0^1 dx \int_{-\infty}^{\infty} dy \
\log^2\left(1+\frac{2}{\sec (\pi  x) \cosh (\pi y)-1}\right) \ \sin^2 (\pi x) + \dots \nonumber \\
&=& 0.290713 \ \phi^2 + O(\phi^4)
\end{eqnarray}

Note that there is no contribution of order $\phi$ since the terms in $\sigma$ corresponding to odd powers of $\phi$
are odd with respect to the change  $x \rightarrow 1-x$. As a consequence of this behavior
the second order term in $\sigma$ in the perturbative expansion provides a leading contribution of order
$\phi^4$ (this is consistent with the leading behavior  found in refs~\cite{Avishai91, Granot02}):
\begin{eqnarray}
E_2 &=& - \pi^2 \Delta_2^2 \approx -0.834119 \phi ^4
\end{eqnarray}

As a consequence of this, in order to evaluate the  dominant contribution to the energy of the
fundamental mode, for $\phi \rightarrow 0$, one needs to take into account also the
contributions of order $\phi^4$ arising from the third and fourth order perturbative
expansion in $\sigma$ (the second order contribution accounts only for about $40 \%$ of the
energy).

We need to consider the third order contribution, calculated in \cite{Amore15}:
\begin{eqnarray}
E_0^{(3)} &=& - 2 \pi^2 \Delta_2 (\Lambda_1 - \Delta_3)
\label{eq_third_order}
\end{eqnarray}

Using the properties of the component of $\sigma$ of order $\phi$ under the change $x \rightarrow 1-x$,
it is easy to see that
\begin{eqnarray}
\Delta_3 =  \pi^3 \int dx_1 dy_1 \int dx_2 dy_2 |y_1-y_2| \sigma(x_1,y_1)\sigma(x_2,y_2) \sin^2 (\pi x_1) \sin^2(\pi x_2) = O(\phi^4)
\end{eqnarray}
and therefore we can neglect this term in the expression for $E_0^{(3)}$.

The remaining expression contains the integral
\begin{eqnarray}
\Lambda_1 &\equiv& \pi^3 \int_{-\infty}^{\infty} dy_1\int_{0}^{1} dx_1\int_{-\infty}^{\infty} dy_2\int_{0}^{1} dx_2
 \sin \left(\pi  x_1\right) \sin \left(\pi  x_2\right)
 \sigma \left(x_1,y_1\right) \sigma\left(x_2,y_2\right)  {G}_2^{(0)}\left(x_1,y_1,x_2,y_2\right) \nonumber \\
&=&   \pi^3  \frac{4 \phi^2}{\pi^2}\int_{-\infty}^{\infty} dy_1\int_{0}^{1} dx_1\int_{-\infty}^{\infty} dy_2\int_{0}^{1} dx_2   \sin \left(\pi  x_1\right) \sin \left(\pi  x_2\right) {G}_2^{(0,odd)}\left(x_1,y_1,x_2,y_2\right)
\nonumber \\
&\times& \log \left(1 + \frac{2}{\sec (\pi  x_1) \cosh (\pi y_1)-1}\right)
\log \left(1 + \frac{2}{\sec (\pi  x_2) \cosh (\pi y_2)-1}\right)  + O(\phi^3)  \nonumber \\
&\approx&  0.171407 \phi^2
\end{eqnarray}
and the third order correction to the energy reads
\begin{eqnarray}
E_3 &=& - 2 \pi^2 \Delta_2 \Lambda_1 + O(\phi^5) \approx -0.983607 \phi ^4
\end{eqnarray}
providing approximately $47 \%$ to the leading order correction.

The calculation of the fourth order contribution requires selecting in the expression for $E_0^{(4)}$
the contributions of order $\phi^4$, using the explicit expressions for $\Delta_i$ and $\Lambda_i$ given above.

A simple inspection proves that there is only one term contributing to the leading order, and the energy reduces to
\begin{eqnarray}
E_4 &=& - \pi^2 \Lambda_1^2 \approx -0.289971 \phi ^4
\end{eqnarray}
corresponding to roughly the $13 \%$ of the total correction.

When we join the three contributions we find
\begin{eqnarray}
\Delta E \approx - \pi^2  \left[ \Delta_2 + \Lambda_1 \right]^2 \approx -2.1077 \phi ^4
\end{eqnarray}
that is extremely close to the value of $c_b$ estimated by Avishai et al.~\cite{Avishai91}.

In Table \ref{tab:table1} we compare this theoretical value with the values obtained numerically,
following the approach of Granot~ref.\cite{Granot02} for $\phi = \pi/10^4$, using an increasing
number of points $n_b$ at which the continuity of the solution is imposed.

In  Fig. \ref{Figure_5} we plot the values in the table and compare the asymptotic coefficient
obtained from the  best fit of these values with the theoretical values obtained from the explicit formulas.
The best fit of the numerical values shows an excellent agreement with the theoretical value calculated
using the perturbative formulas of Refs.\cite{Amore15,Amore16}.

\begin{table}
\caption{\label{tab:table1} Asymptotic coefficient $c_b$ estimated for $\phi = \pi/10^4$ with
$n_b$ border points. }
\begin{center}
\begin{tabular}{|c|c|c|c}
\hline
$n_b$  & $c_b$  \\
\hline
100 & 2.101101624 \\
200 &  2.105792895 \\
300 &  2.10677075 \\
400 &  2.107135377 \\
500 &  2.10731157 \\
600 &  2.107410434\\
700 &  2.107471611\\
800 &  2.10751218 \\
900 &  2.107540509 \\
1000 & 2.107561099 \\
\hline
theoretical & 2.1077 \\
\hline
\end{tabular}
\end{center}
\end{table}

\section{Conclusions}
\label{conclusions}

We have considered three examples of infinite waveguides where the exact perturbative formulas obtained in refs.~\cite{Amore15,Amore16}
apply. In particular for the cases of an infinite homogeneous and asymmetric waveguide and of a broken strip our results agree with
the analogous results obtained applying a formula derived by Bulla et al.~\cite{Bulla97} (asymmetric waveguide) and with the
numerical results obtained in Refs.~\cite{Avishai91, Granot02} (broken strip). Note that the formula of \cite{Bulla97} is limited to
the case of asymmetric homogeneous waveguides, whereas the formulas of Refs.~\cite{Amore15,Amore16} apply to more general geometries (the broken strip is
just one example), even in presence of heterogeneities. This last case has been studied in the second example, an asymmetric heterogeneous waveguide,
showing that the theoretical results obtained using the formulas in \cite{Amore15} are in perfect agreement with the numerical results
obtained using a collocation scheme. To the best of our knowledge, Refs.~\cite{Amore15,Amore16} are the only calculation in which the effect
of both deformations and heterogeneity has been derived: this paper provides a numerical verification of those formulas.

\section*{Acknowledgements}
The research of Paolo Amore and Petr Zhevandrov was supported by the Sistema Nacional de Investigadores (M\'exico).
J. P. Boyd was supported by the National Science Foundation of the U. S. under DMS-1521158.
The figures were produced using  Tikz \cite{tikz}.


\end{document}